\documentclass[12pt,preprint]{aastex}
\usepackage{epsfig}
\usepackage{graphicx}
\usepackage{color}
\usepackage{enumerate,amsmath}
\newcommand{\kms}{{\rm km\,s^{-1}}}
\newcommand{\beq}{\begin{equation}}
\newcommand{\eeq}{\end{equation}}
\newcommand{\ba}{\begin{eqnarray}}
\newcommand{\ea}{\end{eqnarray}}

\def\msun{M_\odot}
\def\rsun{R_\odot}
\def\mbh{M_\bullet}
\def\mimbh{M_{\rm IMBH}}
\def\jlc{J_{\rm lc}}

\def\rmd{d}
\def\peryr{{\rm yr^{-1}}}
\def\E{{\cal E}}
\def\F{{\cal F}}

\lefthead{Xian Chen \& F. K. Liu}
\righthead{An intermediate massive black hole in the Galactic center?}

\begin{document}
\title{Is there an intermediate massive black hole in the Galactic center: Imprints on the stellar tidal-disruption rate}
 
\author{Xian Chen\altaffilmark{1}, \& F. K. Liu\altaffilmark{2}}

\altaffiltext{1}{Kavli Institute for Astronomy and Astrophysics, Peking University, 100871 Beijing, China; {\it chenxian@pku.edu.cn}}
\altaffiltext{2}{Department of Astronomy, Peking University, 100871 Beijing, China; {\it fkliu@pku.edu.cn}}

\begin{abstract}

It has been suggested that an intermediate-massive black hole (IMBH)
with mass $10^{3-5}~M_\odot$ could fall into the galactic center (GC)
and form a massive black hole binary (MBHB) with the central
supermassive black hole, but current observations are not sensitive
to constrain all mass and distance ranges. Motivated by the recent
discovery that MBHBs could enhance the rate of tidal-disruption events
(TDEs) of stellar objects, we investigate the prospect of using
stellar-disruption rate to probe IMBHs in the GC. We incorporated the
perturbation by an IMBH into the loss-cone theory and calculated the
stellar-disruption rates in the GC. We found that an IMBH heavier than
$2000~M_\odot$ could distinguishably enhance the stellar-disruption
rate. By comparing observations of Sgr A* with the fall-back model for
stellar debris, we suggested that the TDE rate in our Galaxy should
not significantly exceed $0.002\, {\rm yr^{-1}}$, therefore a fraction of the
parameter space for the IMBH, concentrating at the high-mass end, can
already be excluded. To derive constraint in the remaining parameter
space, it is crucial to observationally confirm or reject the
stellar-disruption rate between $10^{-4}$ and $10^{-2}~{\rm yr^{-1}}$,
and we discussed possible strategies to make such measurements. 

\end{abstract}


\section{Introduction}\label{sec:intro}

It has been suggested that in the center of our own Galaxy, besides an
unambiguously detected supermassive black hole (SMBH) of
$4.3\times10^6~\msun$ whose location coincides with Sgr A*
\citep[][and references therein]{gen10}, there might be another
intermediate massive black hole (IMBH), which is brought in by an
infalling satellite galaxy \citep{lang11} or core-collapse star
cluster \citep{por06}. 
The existence of massive black hole binary (MBHB) in the Galactic
center (GC) would ease the tension between stellar dynamics and some
observations, e.g. the formation of a young stellar disk inside
$10\arcsec$ (about 4 pc), the central hole of size $5\arcsec$
discovered in the old stellar population, the apparent high velocity
dispersion of a stellar association IRS 13E located at $3.5\arcsec$
from Sgr A*, and the random orientation of the S-stars inside
$1\arcsec$  \citep[see][for a review]{gen10}.  

However, there has been no conclusive observational evidence for or
against an IMBH at the galactic center. Based on the proximity of Sgr
A* to the center of the Galactic nuclear cluster \citep{yu03}, and
more exclusively on the proper motion of Sgr A* \citep{han03,rei04},
most parameter space can be excluded for IMBHs of masses
$\mimbh\ga2\times10^3~\msun$ residing at distances $d>2$ milliparsec
(mpc) from Sgr A*. Slight improvement on the constraint can be
obtained by taking into account the orbital stability of S-stars,
which mostly excludes IMBHs of masses $\mimbh\ga10^5~\msun$ at
$0.3~{\rm mpc}\la d\la10~{\rm mpc}$ \citep{yu03}, and
$2\times10^3~\msun\la\mimbh\la4\times10^3~\msun$ at $2~{\rm mpc}\la
d\la10~{\rm mpc}$ \citep{gua09}. Besides, the short coalescence
timescale ($\la 10^7 \, {\rm yr}$) due to gravitational wave radiation also
argues against IMBHs of any mass within a distance of $d\sim0.1$ mpc
from Sgr A* \citep{han03}. Nevertheless, not all parameter space is
excluded \citep[e.g. see Figure~13 in][]{gua09}.  The low-mass end at
$\mimbh<2\times10^3~\msun$ is currently unaccessible, except that the
IMBH should have $d>0.1$ mpc to comply with the argument of
coalescence timescale. Even in the high-mass end, it is not ruled out
that an IMBH with $2\times10^3~\msun\la\mimbh\la4\times10^4~\msun$ may
reside at either $0.1~{\rm mpc}\la d\la2~{\rm mpc}$ or
$0.06(\mimbh/10^4\msun)^2~{\rm
  pc}\la\mimbh\la3.2(\mimbh/10^4\msun)~{\rm pc}$. 

Previous studies on the stellar orbits bound to MBHBs discovered that the three-body interactions could significantly boost the orbital eccentricities of the stars, partially due to the secular Lidov-Kozai effect \citep{ivanov05} and more importantly due to  chaotic interactions \citep{chen09}. Consequently, compared to that in the single-black-hole case, a larger fraction of stars could reach the so called ``tidal radius'' of the central SMBH to be tidally disrupted, resulting in a burst of stellar-disruption events \citep{chen11,wegg11}. A stellar tidal-disruption event (TDE) will produce a powerful flare by releasing the gravitational energy of the stellar debris \citep{rees88,ulmer99} and is observable in multiple electromagnetic bands ranging from radio to $\gamma$-ray \citep{kom02,gez09,van11,blo11,bur11}. Therefore, if IMBHs indeed reside in the GC, they should as well enhance the rate of TDEs, which may be imprinted in the emissions from the GC.

In this paper we first study the stellar-disruption rate in the GC in
the presence of an IMBH. Based on our results, we discuss the prospect
of constraining the parameters of the hypothetical IMBH using the
observational TDE rates. The outline of the paper is as follows. In
Section~\ref{sec:theory} we develop the loss-cone theory to incorporate the
perturbation by an IMBH, then in Section~\ref{sec:results} we calculate the
stellar-disruption rate in the GC including the effects of the
IMBH. In Section~\ref{sec:obs}, we investigate the observability of a TDE
from GC based on the standard fall-back model for the stellar
debris. Finally, we discuss our results in Section~\ref{sec:discuss}. 

\section{Loss-cone theory}\label{sec:theory}

Around the SMBH in our own Galaxy, the stars with pericenter distances smaller than
\begin{eqnarray}\label{eqn:rt}
r_{t}&\simeq&r_*\left(\frac{M_\bullet}{m_*}\right)^{1/3}\\
&\simeq& 10^{-5.4}~{\rm pc}
\left(\frac{\mbh}{10^{6.6}\msun}\right)^{1/3}\left(\frac{r_*}{\rsun}\right)\left(\frac{\msun}{m_*}\right)^{1/3},
\end{eqnarray}
are subject to tidal disruption \citep{hills75,rees88}, where $r_*$
and $m_*$ are, respectively, the radius and mass of star, and $\rsun$
and $\msun$ refer to the solar values. For $m_*=\msun$ and
$r_*=\rsun$, this ``tidal radius'' is about 10 times greater than the
Schwarzschild radius of the Galactic SMBH and has an angular size of
$0.1$ mas when viewed from the Earth. Tidal disruption creates a
stellar-deficient region in the phase space of specific binding energy
$\E$ and specific angular momentum ${\bf J}$ of stars. This region
is conventionally referred to as the ``loss cone'' because of its
cone-like geometry when the system is spherically symmetric
\citep{frank76,lightman77}. In the following, we restrict our
calculations to the stars bound to the SMBH and far from the tidal
radius, because they dominate the tidal-disruption rate. This
corresponds to an energy range of $\sigma^2\la\E\ll G\mbh/r_t$, where
$\sigma\simeq75~\kms$ is the one-dimensional stellar velocity dispersion in the GC
\citep{gen10}. These stars are orbiting the SMBH on near-Keplarian
orbits with semi-major axis $a=G\mbh/2\E$, and the greatest semi-major
axis we consider is $a_{\rm max}=G\mbh/2\sigma^2\simeq1.6~{\rm
  pc}$. Given $\E$, the maximum angular momentum is
$J_c=G\mbh/(2\E)^{1/2}$ and the boundary of the loss cone by
definition is at $\jlc\simeq(2G\mbh r_t)^{1/2}$. 

The rate of tidal disruption is determined by the rate of stars
diffusing into the loss cone, which is normally more efficient in the
$J$ direction of the $\E-J$ phase space ($J=|{\bf J}|$). In the
simplest case where the system has a single stellar population with
mass $m_*$, suppose the successive mutual scattering between a star
with energy $\E$ and the background stars on average induces an
angular-momentum change $J_D(\E)$ during one stellar orbital period
$P(\E)=2\pi G\mbh/(2\E)^{3/2}\simeq1.6\times10^3(a/0.1~{\rm
  pc})^{3/2}~{\rm yr}$, then the ``two-body'' relaxation timescale can
be calculated with $T_{\rm 2b}(\E)\sim PJ_c^2/J_D^2$. The dependence
of $T_{\rm 2b}$ on the square of angular momentum reflects that the
two-body scattering is a random process, so $J_D$ sums up
incoherently. Given the number of stars $n(\E)\rmd\E$ in the energy
range $\E\sim\E+\rmd\E$,  the loss-cone filling rate is proportional
to $n(\E)\rmd\E/T_{\rm 2b}(\E)$. More careful analysis, taking into
account the detailed distribution function $f(\E,J)$ at the loss-cone
boundary and the fact that stars can be deflected into and out of the
loss cone during one $P(\E)$ when $J_D\gg \jlc$, gives the following
form for the loss-cone filling rate due to two-body relaxation: 
\ba
\F_{\rm 2b}(\E)\rmd\E&=&{j_D^2n(\E)\rmd\E / P(\E)},\label{eqn:f2b}
\ea
where
\ba
j_D^2\equiv\min[\jlc^2/J_c^2,(J_D/J_c)^2/\ln(\jlc/J_c)]\label{eqn:theta2}
\ea
\citep{young77,perets07}. To calculate $(J_D/\jlc)^2$, we adopted
Equation~(14d) in \citet{bahcall77}. 

For stars within about $0.1$ pc from the Galactic SMBH, another efficient relaxation process is resonant relaxation (RR), a coherent change of $J$ driven by the torque exerted by the grainy gravitational field of finite number of stars \citep{rau96,rau98}. Our scheme to calculate the loss-cone filling rate due to RR is analogous to that in \citet{hop06}, but we also revised their formula to include the dependence of RR on stellar orbital eccentricity $e=[1-(J/J_c)^2]^{1/2}$ \citep{gur07}. In the standard case with a single black hole (BH), the timescale for coherent variation of $J$ by RR is limited by the orbital precession timescales induced by general relativity (GR) and by the non-Keplerian potential of the surrounding stellar cusp. Given $P(\E)$ and $\epsilon=(1-e^2)^{1/2}$ for a stellar orbit, we calculate $t_{\rm GR}={(2\epsilon^2/3)}{(a/r_S)}P(\E)$ for the GR precession timescale and $t_M\simeq{\epsilon^{-1}(\mbh/m_*)}{P(\E)/N(>\E)}$ for the cusp-induced precession, where $r_S=2G\mbh/c^2$ is the Schwarzschild radius and $N(>\E)$ denotes the number of stars with energy greater than $\E$. Then the coherent variation timescale, $t_\omega$, can be derived from $1/t_\omega=|1/t_{\rm GR}-1/t_M|$, the minus sign before $1/t_M$ due to the opposite precession directions. During $t_\omega$, $J$ varies coherently by $\Delta J_\omega\simeq\dot{J}_{\rm RR}t_\omega$, where $\dot{J}_{\rm RR}\simeq 0.25eN^{1/2}(>\E)Gm_*/a$ is the RR torque exerted on the stellar orbit \citep{gur07}. On longer timescale $t\gg t_\omega$, the coherence is broken due to orbital precession, therefore $J$ varies incoherently as $\Delta J\simeq \Delta J_\omega(t/t_\omega)^{1/2}$. The RR timescale to erase the initial angular momentum, $J=\epsilon J_c$, is then $T_{\rm RR}\sim (\epsilon J_c/\Delta J_\omega)^2t_\omega$, and more precisely
\begin{eqnarray}
T_{\rm RR}(\E,J)&\simeq&\frac{2.55\epsilon^2}{e^2}\frac{P^2(\E)}{N(>\E)}\left(\frac{\mbh}{m_*}\right)^2\frac{1}{t_\omega}
\end{eqnarray}
\citep{gur07}. The average relaxation timescale $\bar{T}_{\rm RR}$ for the stars with the same energy $\E$ is given by the integration of the equation $\rmd J^2/J_c^2=\rmd t/T_{\rm RR}(\E,J)$, i.e.,
\begin{eqnarray}
\bar{T}_{\rm RR}(\E)&=&\int_{\jlc^2}^{J_c^2}T_{\rm RR}(\E,J)\rmd J^2/{J_c^2}.\label{eqn:trrbar}
\end{eqnarray}
Then loss-cone filling rate due to RR can be calculated with
\ba
\F_{\rm RR}(\E)\rmd\E&=&{n(\E)\rmd\E / \bar{T}_{\rm RR}(\E)}.
\ea

If an IMBH with mass $\mimbh$ (the mass ratio of the IMBH-SMBH binary being $q\equiv\mimbh/\mbh$) resides at a distance $d$ from Sgr A*, the tidal force of the IMBH also exerts a torque on a stellar orbit, which induces an additional coherent variation of $J$ at the rate $\dot{J}_K\simeq J_c/T_K$, where
\begin{eqnarray}
T_{K}&=&
\left\{
\begin{array}{ll}
\frac{2}{3\pi q}\left(\frac{a}{d}\right)^{-3}P(a)\,\,\,\,\,\,\,\,(a\le d/2)\\
\frac{16\sqrt{2}}{3\pi q}\left(\frac{a}{d}\right)^{1/2}P(a)\,\,\,\,\,\,\,\,(a> d/2)
\end{array}
\right.
\end{eqnarray}
is analogous to the Lidov-Kozai timescale \citep{lidov62,kozai62} but
also accounts for the stars in the chaotic regime with $a\sim d$ which
dominates the loss-cone refilling \citep{chen09,chen11}. We note that
during chaotic interactions the stellar orbits could reach extreme
eccentricities ($e\simeq1$) irrespective of their initial inclinations
or the $z$-components of angular momenta \citep{chen11}, which are
fundamentally different from those interactions in the secular
Lidov-Kozai mechanism and significantly increase the stellar
reservoir for tidal disruption. Because of the extra nodal precession
induced on the stellar orbit by the IMBH, the coherent variation of
$J$ is limited by the new timescale $t'_\omega=|1/t_K+1/t_{\rm
  GR}-1/t_M|^{-1}$, where $t_K\simeq\epsilon T_K$
\citep{ivanov05}. During $t'_\omega$, the variation of $J^2$, due to
both RR and the IMBH perturber, is $\Delta
(J'_\omega)^2=(\dot{J}^2_{\rm RR}+\dot{J}^2_K)(t'_\omega)^2$. The
resulting timescale for erasing the initial $J$ is $T_{\rm
  co}(\E,J)\simeq t'_\omega (\epsilon J_c)^2/\Delta
(J'_\omega)^2$. Therefore, the coherent relaxation timescale due to
both RR and an IMBH is 
\begin{eqnarray}
\bar{T}_{\rm co}(\E)&=&\int_{\jlc^2}^{J_c^2}T_{\rm co}(\E,J)\rmd J^2/{J_c^2},\label{eqn:tcobar}
\end{eqnarray}
and the corresponding loss-cone filling rate is
\ba
\F_{\rm co}(\E)\rmd\E&=&{(1-f_{\rm ej})n(\E)\rmd\E / \bar{T}_{\rm co}(\E)}.\label{eqn:fco}
\ea
We assume $f_{\rm ej}\simeq0.5$ to correct, to the zeroth order, the star loss due to slingshot ejection, since $f_{\rm ej}$ is a complex function of $q$, $a/d$, and the orbital eccentricity of the IMBH \citep{chen11}. 

\section{Stellar-disruption rate in the GC}\label{sec:results}

To give $n(\E)\rmd\E$ and $N(>\E)$, we adopted a stellar distribution
for the GC from \citet{schodel07}, whose mass volume density has an
outer Bachall-Wolf and an inner $\gamma=1.2$ power-law profile with a
break radius at $r_b\simeq0.22$ pc. The corresponding stellar
distribution function $f(\E)$ scales as $f(\E)\propto\E^{-p}$, where
$p=1/4~(-0.3)$ for $\E\ll G\mbh/r_b~(\E\gg G\mbh/r_b)$. The
normalization for $f(\E)$ is derived from $\rho(r)=4\pi
m_*\int_0^{G\mbh/r}\sqrt{2(G\mbh/r-\E)}f(\E)\rmd\E$, then by
definition we derive $n(\E)=4\pi^2J_c^2f(\E)P(\E)$ and
$N(\E)=\int_{\E}^{G\mbh/2r_t} n(\E)\rmd\E$. Using these quantities, as
well as Equations~(\ref{eqn:fco}) and (\ref{eqn:f2b}), we calculated
the stellar-disruption rates in the GC for a grid of hypothetical
IMBHs of different $\mimbh$ and $d$. 

\begin{figure}
\plotone{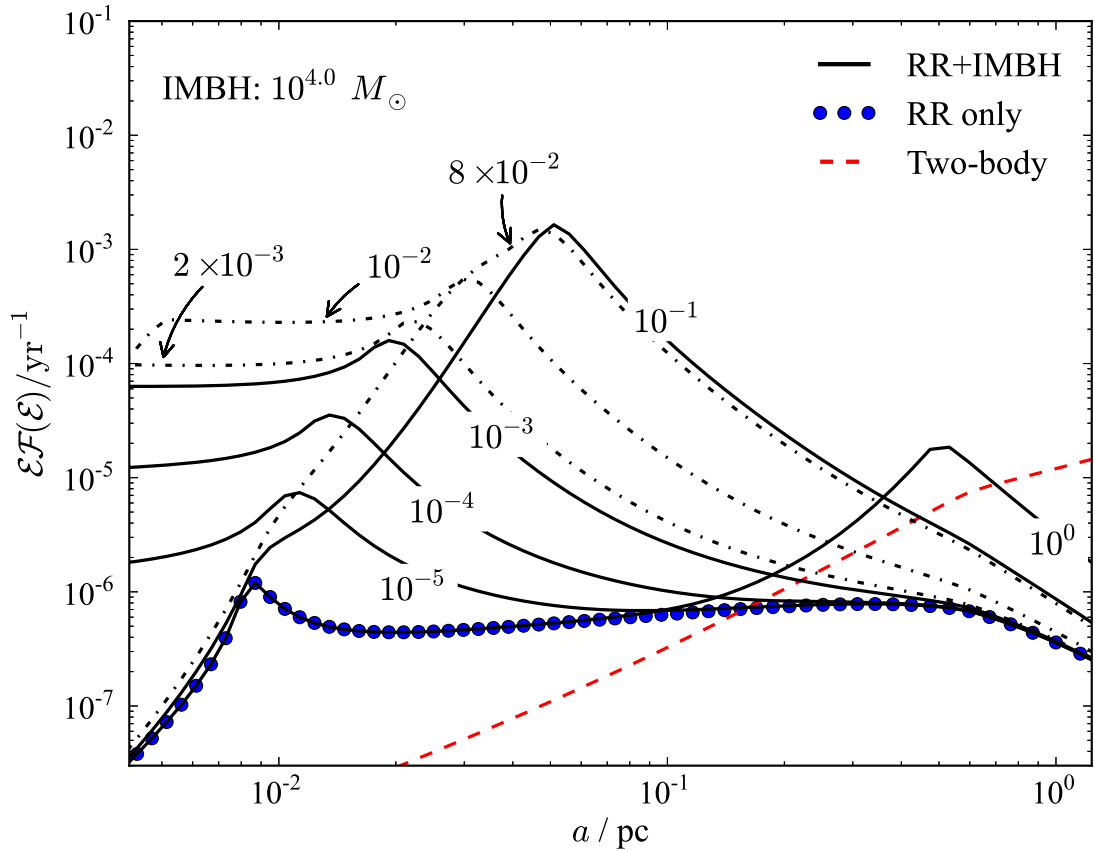}
   \caption{Contribution to loss-cone filling rate by stars at
   different semi-major axis, assuming $m_*=\msun$. Different
   stellar relaxation mechanisms are indicated with different line
   styles, i.e., dashed for two-body relaxation, dotted for RR only,
   and solid for RR+IMBH with $\mimbh=10^4~\msun$. The annotations on
   the lines are the assumed 
   distances ($d$) in unit of pc for the IMBH. An IMBH with
   $10^4~\msun$ at $d=(2\times10^{-3},\,10^{-2},\,8\times10^{-2})$ pc
   is excluded by observations and thus the loss-cone filling rates
   due to the mechanisms RR+IMBH are shown in dot-dashed lines.}
   \label{fig:m4}
\end{figure}

The dashed line in Figure~\ref{fig:m4} shows the stellar-disruption
rate contributed by two-body relaxation at different semi-major axis,
assuming $m_*=1~\msun$. Since $J_D\propto m_*^{1/2}$ and $r_*\propto
m_*^{0.8}$ for main-sequence stars when $m_*\la20~\msun$ \citep{kw90},
the rate $\F_{\rm 2b}$ scales only mildly with $m_*$, as $1/\ln m_*$
in the limit $J_D\ll J_{\rm lc}$ and $m_*^{0.53}$ in the limit $J_D\gg
J_{\rm lc}$ (see Equation~(\ref{eqn:theta2})). We do not show the
contribution from the stars at $a>1.6$ pc, because the Keplerian
assumption would break down for these stars, but their contribution to
the integrated loss-cone filling rate $\dot{N}=\int\F(\E)\rmd\E$ is
insignificant anyway. Inside the central 0.004 pc, the stellar mass is
dominated by compact objects such as white dwarfs and stellar-mass
black holes \citep{hop06b}, therefore the rates inside $0.004$ pc are
not shown either. The total stellar-disruption rate due to incoherent
relaxation, integrated over the range $0.004~{\rm pc}<a<1.6~{\rm pc}$
is $\dot{N}_{\rm 2b}\simeq3.5\times10^{-5}~\peryr$, agrees well with
the previous more sophisticated calculations \citep[e.g.][]{mer10}.  

The dotted line in Figure~\ref{fig:m4} shows the contribution to stellar-disruption rate by RR only, without the perturbation of an IMBH. Compared to two-body relaxation (dashed line), RR becomes more important at $a<0.1$ pc \citep[also see][]{hop06}, and the sharp decline of $\F(\E)\E$ inside $a\simeq 0.01$ pc is caused by the quenching of RR by relativistic precession. The integrated stellar-disruption rate is $\dot{N}_{\rm RR}\simeq3.5\times10^{-6}~\peryr$, and it does not depend on $m_*$ because both $n(\E)\rmd\E$ and $\bar{T}_{\rm RR}(\E)$ scales as $m_*^{-1}$.

If an IMBH with $\mimbh=10^4~\msun$ resides in the GC, the total
stellar-disruption rates contributed by the coherent relaxations,
i.e. RR plus IMBH perturbation, are shown in Figure~\ref{fig:m4} as
the solid and dot-dashed lines. The rates are derived under the
assumption $m_*=\msun$ and scale 
as $m_*^{-1}$ if $m_*$ varies. Comparing the solid lines with the
dotted one, we found that the presence of the IMBH dramatically
enhances the loss-cone filling rate. The integrated stellar-disruption
rate derived form the solid lines, $\dot{N}_{\rm co}$, increases with
decreasing $d$ when $d\ga0.1$ pc. It reaches a maximum of
$\dot{N}_{\rm co}\simeq1.1\times10^{-3}~\peryr$ when $d\simeq0.07$ pc,
which is a factor of $300$ greater than the unperturbed value of
$\dot{N}_{\rm RR}$, and $30$ times higher than $\dot{N}_{\rm 2b}$. As
$d$ further decreases from $d=0.07$ pc, the integrated rate
$\dot{N}_{\rm co}$ becomes smaller, because the coherent variation of
$J$ at $a\sim d$ becomes more susceptible to quench by relativistic
precession. 

\begin{figure}
\plotone{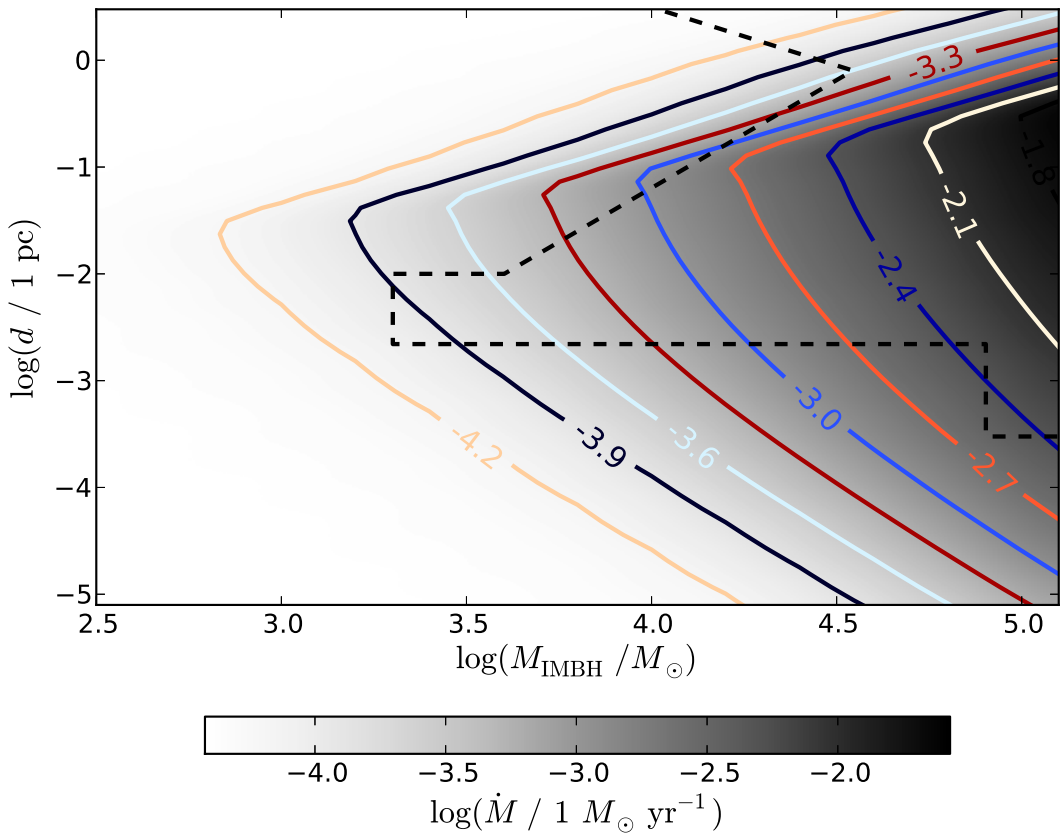}
   \caption{Integrated mass disruption rates (intensity map and
     contours), including contributions from both coherent and
     incoherent relaxations, for a grid of $M_{\rm IMBH}$ and
     $d$. The parameter space for IMBH excluded by the dynamics
     of S-stars and Sgr A* \citep{gua09} is the region to the right
     hand side of the dashed line.}
   \label{fig:con}
\end{figure}

Figure~\ref{fig:con} (intensity map and contours) shows the total mass
disruption rate,  $\dot{M}=m_*(\dot{N}_{\rm 2b}+\dot{N}_{\rm co})$,
which is insensitive to the assumption of $m_*$, as a function of
$\mimbh$ and $d$. The dashed line indicates the parameter space that
is excluded by the observed dynamics of S-stars and Sgr A*
\citep{gua09}. In general, $\dot{M}$ is a increasing function of
$\mimbh$ (because $\dot{N}_{\rm co}\propto\mimbh$), and given
$\mimbh$, the rate  peaks at $d\sim0.01-0.5~{\rm pc}$. For example,
when $\mimbh=(10^3,10^4,10^5)~\msun$, the maxima of $\dot{M}$ are
$(8.3,110,1600)\times10^{-5}~\msun~\peryr$, occurring at
$d=(0.03,0.07,0.3)~{\rm pc}$. To the left-hand-side of the dashed
line, where the possibility of an IMBH cannot be excluded by the
current observations, the perturbed stellar-disruption rate ranges
from $3\times10^{-5}~\msun~\peryr$ when $\mimbh\le10^3~\msun$ to as
high as $4\times10^{-3}~\msun~\peryr$ when
$\mimbh\simeq10^5~\msun$. We also derived $\dot{N}_{\rm co}$ for
$q=1/81$, $1/243$, and $1/729$ from the data of our previous
scattering experiments in pseudo-Newtonian gravitational potentials
\citep{chen11}. The resulting numerical rates agree with those from
the above analytical calculations within a factor of $2$, despite many
simplifications in our analytical model. 

\section{Observability of TDEs in the GC}\label{sec:obs}

The black contour in Figure~\ref{fig:con}
($\dot{M}\simeq10^{-4}~\msun~\peryr$) indicates that an IMBH with
$\mimbh>2000~\msun$, if currently resides in the GC, would
significantly raise the possibility of TDEs in the past $10^2$--$10^4$
years. The standard model for TDE \citep{rees88,ulmer99} predicts that
a fraction of the stellar debris will ``fall back'' to the periastron,
forming an accretion disk and producing a shock-heated hot spot, whose
initial luminosity should be close to or even exceeding the Eddington
luminosity of the central SMBH ($L_E\simeq5.5\times10^{44}~{\rm
  erg~s^{-1}}$ for GC). Is the hypothesized high stellar-disruption
rate consistent with the observed quiescent state of Sgr A*?  

In the framework of the ``fall back model'', the most bound seller
debris has a binding energy of  $\Delta\E\simeq kG\mbh
r_*/(r_t/\beta)^2$ and will return to the periastron after a time
delay of 
\ba
t_{\rm min}&=& 2\pi G\mbh(2\Delta\E)^{-3/2}\\
&\simeq&0.22k^{-3/2}\beta^{-3}(m_*/\msun)^{-1}(r_*/\rsun)^{3/2}~{\rm yr}
\ea
\citep{li02}, where $k\simeq 1 -- 3$ accounts for the stellar spin at
periastron before tidal disruption and $\beta\equiv r_t/r_p\sim1$
denotes the ratio between the tidal radius $r_t$ and the pericenter
distance $r_p$. About a fraction $f\sim0.5$ of the stellar debris
remains bound to the central SMBH \citep{eva89}, and the resulting
fall-back rate as a function of time is 
\ba
\dot{M}_{\rm fb}&=&\frac{2fm_*}{3t_{\rm min}}\left(\frac{t-t_D}{t_{\rm min}}\right)^{-5/3}
\simeq0.12~{\msun~\peryr}~k^{-1}\beta^{-2}\nonumber\\
&\times&\left(f\over0.5\right)\left(m_*\over\msun\right)^{1/3}\left(r_*\over\rsun\right)\left(\frac{t-t_D}{1~{\rm
    yr}}\right)^{-5/3},\label{eq:mdot} 
\ea
where $t_D$ denotes the time of stellar disruption and $t-t_D\ge t_{\rm min}$. 
We note that the fall-back rate when $\mimbh>4\times10^4~\msun$ and
$d>0.003$ pc is not calculated self-consistently, because in this
case, according to \citet{liu09}, $\dot{M}_{\rm fb}$ would be
interrupted at $t-t_D>10$~yr. Anyway, this parameter space is
excluded by the dynamics of Sgr A* and S-stars \citep{gua09}. Because
of mutual collisions when returning to the periastron, the fall-back
material circularizes at a radius of about $2r_p$, and the prompt
release of the kinetic energy gives rise to a hot spot with bolometric
luminosity 
\ba
L_{\rm hs}&=&\varepsilon\dot{M}_{\rm fb}c^2\simeq9.5\times10^{42}~{\rm erg~s^{-1}}\nonumber\\
&\times&k^{-1}\beta^{-1}\left(f\over0.5\right)\left(m_*\over\msun\right)^{2/3}\left(\frac{t-t_D}{1~{\rm yr}}\right)^{-5/3},\label{eq:lbol}
\ea
where $\varepsilon\simeq G\mbh/(4r_pc^2)$ is the conversion efficiency
of kinetic energy into radiation and $c$ is the speed of light. The
spectral energy distribution (SED) of the hot spot is mostly likely
black-body (BB) because of the large opacity of the fall-back
material. The effective temperature of the BB depends on the emission
area, $A_{\rm BB}$, which is uncertain due to the current poor
understanding of the structures of the colliding streams
\citep{koc94,kim99}. Since the vertical scale hight orthogonal to the
orbital plane of the stellar stream is greater than $r_*$, the cross
section for stream collision should not be much smaller than $2\pi
r_*r_p$, and the geometry factor $\zeta=A_{\rm BB}/(4\pi r_p^2)$ for
the emission area is likely greater than $r_*/r_p\sim0.01$. For a
typical value of $\zeta=0.1$, the effective temperature of the hot
spot is 
\ba
T_{\rm BB}&=&\left(\frac{L_{\rm bol}} {\sigma_S A_{\rm BB}}\right)^{1/4}\simeq\left(\beta^2L_{\rm bol}\over4\pi\sigma_S \zeta r_t^2\right)^{1/4}\nonumber\\
&\simeq&1.8\times10^5~{\rm K}~\left(\beta\over
k\right)^{1/4}\left(f\over0.5\right)^{1/4}\left(m_*\over\msun\right)^{1/3}\nonumber\\ 
&\times&\left(\frac{\zeta}{0.1}\right)^{-1/4}\left(r_*\over\rsun\right)^{-1/2}\left(\frac{t-t_D}{1~{\rm yr}}\right)^{-5/12},\label{eqn:tbb}
\ea
where $\sigma_S$ is the Stefan-Boltzmann constant.

\begin{deluxetable}{lllll}
\tabletypesize{}
\tablecolumns{5}
\tablecaption{Predictions from the Fall-back Model\label{tab:pre}}
\tablehead{\colhead{$t-t_D$ (years)}&\colhead{$10$}&\colhead{$10^2$}&\colhead{$10^3$}&\colhead{$10^4$}}
\startdata
$\dot{M}_{\rm fb}$ ($\msun~\peryr$)&$~10^{-2.6}$&$10^{-4.3}$&$10^{-5.9}$&$10^{-7.7}$\\
&$(10^{-4.5}$&$10^{-6.1}$&$10^{-7.8}$&$10^{-9.3}$)\\
$L_{\rm hs}$ (${\rm erg~s^{-1}}$)&$~10^{41.3}$&$10^{39.6}$&$10^{38.0}$&$10^{36.3}$\\
&($10^{40.1}$&$10^{38.5}$&$10^{36.8}$&$10^{35.1}$)\\
$T_{\rm BB}$ ($10^3 $K)&$~79$&31&12&4.5\\
&(160&61&23&8.9)\\
$F_\nu(K)$ (mJy)&$~160$&56&15&2.0\\
&(35&12&4.1&0.96)
\enddata
\end{deluxetable}

To get a sense of the brightness of the hot spot in the GC, we
 calculated $\dot{M}_{\rm fb}$ and $L_{\rm hs}$ as functions of
 $t-t_D$ with the fiducial parameters
 $(f,k,\beta,\zeta,r_*,m_*)=(0.5,1,1,0.1,\rsun,\msun)$. We also
 increased $k$ and $\beta$ to $k=3$ and $\beta=5$ to estimate the
 lower limits to $\dot{M}_{\rm fb}$ and $L_{\rm hs}$. The results are
 given in Table~\ref{tab:pre} with the lower limits bracketed by
 parentheses.  We found that 10 years after a TDE in the GC, the
 fall-back rate has already dropped to about $30$-$2600$ times smaller
 than the Eddington rate, defined as $\dot{M}_E\equiv
 L_E/(0.1c^2)\simeq8.4\times10^{-2}~\msun~\peryr$. By this time, the
 bolometric luminosity of the hot spot is $4-5$ orders of magnitude
 smaller than the Eddington luminosity, when taking into account that
 $\varepsilon$ is much smaller than the typical value of $0.1$ derived
 from the disk accretion theory.  
 We note that when $t-t_D\ga100~{\rm yr}$, the bolometric luminosity
 contributed by the accretion disk produced by the circularized
 material is negligible because the fall-back rate is too low to
 support a radiatively efficient accretion disk
 \citep{can90,men01}. Even in the extreme case that the accretion flow
 can instantly drain the fall-back material \citep{ulmer99}, the
 luminosity of the radiatively inefficient accretion flow is still as
 low as $10^{38}~{\rm erg~s^{-1}}(\dot{M}_{\rm
 fb}/10^{-4}~\msun~\peryr)^2$ \citep{nar95,nar98}. Observations of Sgr
 A* revealed that the bolometric luminosity is of order
 $10^{36-37}~{\rm erg~s^{-1}}$ \citep{gen10}, and the SED can be well
 modeled with a radiatively inefficient disk with an accretion rate of
 $10^{-6}~\msun~\peryr$ \citep{nar98}. Detailed modeling of the
 accretion flow also indicates that the accretion rate inside 20
 gravitational radii (smaller than the typical tidal radius) of the
 central SMBH is even smaller, ranging from $10^{-9}$ to
 $10^{-7}~\msun~\peryr$ \citep[][and references therein]{shc10}. These
 facts, when compared to $\dot{M}_{\rm fb}$ and $L_{\rm hs}$  derived
 from the fall-back model, suggest that TDEs are unlikely to have
 occurred in our own Galaxy within the last $4-5$ centuries. 

We also calculated $T_{\rm BB}$ of the hot spot according to
Equation~(\ref{eqn:tbb}). The results for the fiducial parameters
(data without parentheses) and for $(k,\beta,\zeta)=(3,5,0.01)$ are
also shown in Table~\ref{tab:pre}. According to Wein's displacement
law, the SED of the hot spot initially peaks at the UV band and will shift
toward IR  as $t-t_D$ increases to $10^4$ years. Since IR bands have
relatively low extinctions toward the GC and are commonly used by the
ground-based telescopes to monitor the GC, we calculated the
monochromatic flux of the hot spot at a wavelength of $2.2\mu$m ($K$-band) with an extinction of $A=3$. The results are given in the last two rows of Table~\ref{tab:pre}, where the data in parentheses correspond to non-fiducial parameters with $(k,\beta,\zeta)=(3,5,0.01)$.
For other non-fiducial combinations of $(k,\beta,\zeta,f,m_*,r_*)$, one can use the scaling $F_\nu\propto L_{\rm bol}/T_{\rm eff}^3$ to derive the IR flux. The $K$-band flux from the radiatively inefficient accretion flow is very difficult to derive because of many unknown model parameters, but is typically smaller than $10(\dot{M}_{\rm fb}/10^{-4}~\msun~\peryr)^2$ mJy \citep{nar98}. Observationally, the $K$-band flux of Sgr A* is measured to be $\la5$ mJy with occasional flares as bright as $30$ mJy \citep[][and references therein]{mor12}. Despite the uncertainties in the model predictions,  we found the null hypothesis that a TDE occurred in the GC within the last $7-8$ centuries to be in severe conflict with the observed $K$-band flux of Sgr A*.

\section{Discussion}\label{sec:discuss}

In order to test the possibility and constrain the parameters of a
hypothetical IMBH in the GC, we studied the effects of an IMBH on the
stellar-disruption rate in the framework of the loss-cone theory. We
found that an IMBH with mass greater than $2000~\msun$ could
significantly enhance the stellar disruption rate in GC
(Figure~\ref{fig:m4}) because it coherently perturbs the angular
momenta of stars in a fashion analogous to but more chaotic than the
Lidov-Kozai effect.  The maximum mass disruption rate increases with
black hole mass as $10^{-3}M_4~\msun~\peryr$, where
$M_4=\mimbh/10^4~\msun$, and is reached when the IMBH arrives at about
$0.1\arcsec\sim1\arcsec$ from Sgr A* (Figure~\ref{fig:con}). These
results have already brought up some intriguing implications. For
example, it is speculated that the high velocity dispersion of IRS
13E, a young stellar association locating at  $0.14$ pc from Sgr A*,
may be induced by an embedding IMBH with mass  greater than
$10^4~\msun$ \citep{mai04,sch05,fri10}. If so, the mass disruption
rate in the GC according to our calculation should be greater than about
$5\times10^{-4}~\msun~\peryr$, up to about $0.02~\msun~\peryr$ in the
extreme case with $\mimbh=10^5~\msun$. The results also imply that an
IMBH in the GC would greatly increase the chance for us to detect
infalling objects on the course toward Sgr A* \citep{gil12}. 

Is the hypothesized high stellar-disruption rate consistent with the
current quiescent state of Sgr A*? To estimate the set-off time for
the most recent TDE in the GC, we calculated the evolution of the
bolometric luminosity and the IR flux of a TDE in the GC using the
fall-back model (Section~\ref{sec:obs}). When comparing the model
predictions with the observations of Sgr A*, we found that the current
quiescent state of Sgr A* only allows a stellar-disruption rate  lower
than about $0.002~\peryr$. According to this limit, the IMBH is
unlikely to reside in the right-hand side of the $\log(\dot{M})=-2.7$
contour in Figure~\ref{fig:con}. Interestingly, most part of this
disfavored parameter space coincides with the space already excluded
by the dynamics of Sgr A* and the S-stars (dashed line in
Figure~\ref{fig:con}), confirming the previous constraints on the
IMBH. In addition, the limit on $\dot{M}$ excludes a patch of the
parameter space at $d\la0.002$ pc and $\dot{M}\ga0.002~\msun~\peryr$,
which was not reached by the dynamics of Sgr A* and the
S-stars. Furthermore, to comply with the limit on $\dot{M}$, the
hypothetical IMBH in IRS 13E should not  exceed $2\times10^4~\msun$. 

The above success encourages us to investigate the prospects of using
stellar-disruption rates to probe even smaller and closer (to Sgr A*)
IMBHs. For the parameter space of IMBH that is still allowed by the
dynamics of Sgr A* and the S-stars (Figure~\ref{fig:con}), the maximum
stellar-disruption rate is $\dot{M}\simeq10^{-3}~\msun~\peryr$ when
$\mimbh=10^4~\msun$ and $\dot{M}\simeq10^{-4}~\msun~\peryr$ when
$\mimbh=10^3~\msun$, and these maxima occur at $d\sim0.04$ pc (about 1
arcsec). Besides, at a distance as close as $d<0.002$ pc (about 50
mas), where there is nearly no constraint on $\mimbh$ by the previous
observations, the maximum $\dot{M}$ varies from about
$4\times10^{-3}~\msun~\peryr$ for $\mimbh=10^5~\msun$ to about
$10^{-4}~\msun~\peryr$ for $\mimbh=2000~\msun$. Combining these
results and taking into account that the stellar-disruption rate for a
single SMBH in the GC is $\dot{N}_{\rm
  2b}\simeq3.5\times10^{-5}~\peryr$, we suggest that a
stellar-disruption rate between $10^{-4}$ and $10^{-2}~\peryr$, if
identified in our own Galaxy, would be a strong evidence for an IMBH
more massive than $2000~\msun$ lying in the central parsec of the
GC. On the other hand, if the stellar-disruption rate higher than
$10^{-4}~\peryr$ is rejected by future observations, a large parameter
space, which is not covered by the current constraints on the
hypothetical IMBH, could be excluded. 

A stellar-disruption rate of $10^{-2}~\peryr$ is already too low to be
measured directly from decadal observational cycles, therefore
measuring a rate down to  $10^{-4}~\peryr$ requires special
strategies. It is observed that the central tens of parsecs of the GC
are populated by molecular clouds \citep[MCs; see][for a
  review]{gen10}.  A TDE at Sgr A* could leave imprints on the MC
system in the form of reflected hard X-rays or fluorescent iron lines
decades or even centuries after the outburst \citep{yu11}. A
systematic search for such ``light echoes'' within 30 pc from Sgr
A* indeed revealed a major flare as luminous as $3\times10^{39}~{\rm
  erg~s^{-1}}$ in X-ray starting to decay at about 150 years ago
\citep[][and references therein]{mor12}. These observations give a
upper limit of $0.004\sim0.006~\peryr$ to the TDE rate in the GC
because according to Equation~(\ref{eq:lbol}) it takes at least
$180-280$ years (depending on $k$ and $\beta$) for the luminosity of
hot spot to drop below $3\times10^{39}~{\rm erg~s^{-1}}$, not to mention that accounting extra X-ray contribution from an accretion disk would extend the time elapse $t-t_D$. To test whether the flare recurs every 200$\sim$300 years, it is crucial to look for a second light echo in the MC system $30-100$ pc away from Sgr A*. On longer timescales ranging from thousands to millions of years, the imprints of TDEs could be detected in the forms of enhanced electron-positron annihilation lines, high-energy gamma rays, and cosmic rays \citep{cheng06,cheng07,cheng11}. If absorbed by the bio-circle, the records of enhanced high-energy emissions could be preserved on the earth surface for thousands of years \citep{miyake12}. In order to constrain $\dot{M}$ in the GC over an extended timescale of $10^{4-6}$ years, we appeal for more efforts in measuring and modeling of the emissions in these high-energy bands.

\acknowledgments

We thank Roman Shcherbakov and Shuo Li for many helpful
discussions. This work is supported by the Chinese national 973
program (2007CB815405) and the National Natural Science Foundation of
China (NSFC11073002). F.K.L. also thanks the support from the China
Scholarship Council (2009601137), and X.C. acknowledges the support from
China Postdoc Science Foundation (2011M500001).

\end{document}